\documentclass[aps,prb,twocolumn,a4paper,showpacs,superscriptaddress,longbibliography]{revtex4-1}

\usepackage{amsmath,amssymb,bm,graphicx}
\usepackage{xcolor}
\usepackage[colorlinks=true,urlcolor=blue,linkcolor=blue,citecolor=blue,bookmarks=false]{hyperref}

\begin{document}

\title{History dependence of the magnetic properties of single-crystal Fe$_{1-x}$Co$_{x}$Si}

\author{A. Bauer}
\email{andreas.bauer@frm2.tum.de}
\affiliation{Physik-Department, Technische Universit\"{a}t M\"{u}nchen, D-85748 Garching, Germany}

\author{M.~Garst}
\affiliation{Institute for Theoretical Physics, Universit\"{a}t zu K\"{o}ln, D-50937 K\"{o}ln, Germany}
\affiliation{Institut f\"{u}r Theoretische Physik, Technische Universit\"{a}t Dresden, D-01062 Dresden, Germany}

\author{C. Pfleiderer}
\affiliation{Physik-Department, Technische Universit\"{a}t M\"{u}nchen, D-85748 Garching, Germany}

\date{\today}

\begin{abstract}
We report the magnetization, ac susceptibility, and specific heat of optically float-zoned single crystals of Fe$_{1-x}$Co$_{x}$Si, $0.20 \leq x \leq 0.50$. We determine the magnetic phase diagrams for all major crystallographic directions and cooling histories. After zero-field cooling, the phase diagrams resemble that of the archetypal stoichiometric cubic chiral magnet MnSi. Besides the helical and conical state, we observe a pocket of skyrmion lattice phase just below the helimagnetic ordering temperature. At the phase boundaries between these states evidence for slow dynamics is observed. When the sample is cooled in small magnetic fields, the phase pocket of skyrmion lattice may persist metastably down to lowest temperatures. Taken together with the large variation of the transition temperatures, transition fields, and the helix wavelength as a function of composition, this hysteresis identifies Fe$_{1-x}$Co$_{x}$Si as an ideal material for future experiments exploring, for instance, the topological unwinding of the skyrmion lattice.
\end{abstract}


\vskip2pc

\maketitle

\section{Motivation}

The discovery of a skyrmion lattice stimulated great scientific interest in chiral magnets crystallizing in space group $P2_{1}3$, which comprises itinerant B20 compounds such as MnSi\cite{2009:Muhlbauer:Science}, Mn$_{1-x}$Fe$_{x}$Si\cite{2010:Pfleiderer:JPhysCondensMatter}, Fe$_{1-x}$Co$_{x}$Si\cite{2010:Munzer:PhysRevB, 2010:Yu:Nature}, and FeGe\cite{2011:Yu:NatureMater} as well as the insulator Cu$_{2}$OSeO$_{3}$\cite{2012:Seki:Science, 2012:Adams:PhysRevLett, 2012:Seki:PhysRevB}. In general, the magnetism in these compounds reflects a well-defined set of three hierarchical energy scales.\cite{1980:Landau:Book} Exchange interactions and Dzyaloshinsky-Moriya spin-orbit interactions on the strongest and intermediate scale generate a long-wavelength helical modulation, while higher-order spin-orbit coupling in zero field aligns the pitch of the helices along certain crystallographic directions. The skyrmion lattice represents a novel form of magnetic order composed of spin vortices with a finite topological winding number. It appears in a phase pocket, long known as the A phase, just below the helimagnetic transition temperature in small magnetic fields.

The skyrmions arrange in a hexagonal lattice in the plane perpendicular to the applied magnetic field and form skyrmion tubes along the field direction. The non-trivial topological winding of the skyrmions gives rise to an emergent electrodynamics, in which each skyrmion carries one quantum of emergent magnetic flux, and allows for an efficient coupling of spin currents to the magnetic texture.\cite{2009:Neubauer:PhysRevLett, 2012:Schulz:NaturePhys} In combination with the exceptionally well-defined long-range order\cite{2011:Adams:PhysRevLett} and the resulting weak collective pinning, this coupling permits to drive the skyrmion lattice at ultra-low current densities.\cite{2010:Jonietz:Science, 2012:Yu:NatCommun} Further ways to control skyrmions include magnon flow induced by thermal gradients\cite{2014:Mochizuki:NatureMater, 2014:Lin:PhysRevLett} or electric fields exploiting the magnetoelectric coupling of the insulating Cu$_{2}$OSeO$_{3}$\cite{2012:Seki:PhysRevB2, 2012:White:JPhysCondensMatter, 2014:White:PhysRevLett}. In addition, a distinct set of collective excitations in the GHz range\cite{2012:Mochizuki:PhysRevLett} may be excited electromagnetically\cite{2012:Onose:PhysRevLett, 2013:Okamura:NatCommun, 2015:Okamura:PhysRevLett, 2015:Schwarze:NatureMater} or optically\cite{2012:Koralek:PhysRevLett, 2015:Ogawa:SciRep} promising new concepts for microwave devices. 

In this context, the pseudo-binary compound Fe$_{1-x}$Co$_{x}$Si is interesting, as it allows to change important parameters drastically by compositional tuning. Helimagnetism is observed in a wide composition range, $0.05 < x < 0.8$\cite{1983:Beille:SolidStateCommun, 1987:Motokawa:JMagnMagnMater, 2004:Manyala:NatureMater}, whereas the parent compounds FeSi and CoSi are paramagnetic\cite{1967:Jaccarino:PhysRev} and diamagnetic\cite{1966:Shinoda:JPhysSocJpn}, respectively. Starting from the strongly correlated insulator FeSi,\cite{2008:Arita:PhysRevB, 2010:Mazurenko:PhysRevB} an insulator-to-metal transition takes place at $x = 0.02$.\cite{1997:Chernikov:PhysRevB, 2007:Zur:PhysRevB, 2008:Manyala:Nature, 2009:Menzel:PhysRevB} Due to the comparably high absolute value of the electrical resistivity and an upturn at low temperatures,\cite{1983:Beille:SolidStateCommun, 1993:Lacerda:PhysicaB, 2004:Manyala:NatureMater, 2005:Onose:PhysRevB} helimagnetic Fe$_{1-x}$Co$_{x}$Si is often referred to as a strongly doped semiconductor. This behavior contrasts studies of both the magnetic properties\cite{1976:Kawarazaki:JPhysSocJpn, 1985:Watanabe:JPhysSocJpn, 1987:Inoue:PhysStatusSolidiB, 1989:Watanabe:JPhysSocJpn, 1992:Ishimoto:JPhysSocJpn, 1998:Ohta:JMagnMagnMater, 2002:Chattopadhyay:PhysRevB, 2002:Chattopadhyay:PhysRevB2} and the band structure\cite{2006:Mena:PhysRevB, 2007:Zur:PhysRevB, 2009:Menzel:PhysRevB, 2010:Mazurenko:PhysRevB} of Fe$_{1-x}$Co$_{x}$Si consistently suggesting itinerant magnetism. 

Depending on the cobalt content, Fe$_{1-x}$Co$_{x}$Si shows helimagnetic transition temperatures between a few Kelvin and 50\,K, while the critical fields vary between a few millitesla and roughly 150\,mT.\cite{1983:Beille:SolidStateCommun, 2005:Onose:PhysRevB, 2007:Grigoriev:PhysRevB2} The helix wavelength ranges from about $300\,\textrm{\AA}$ to more than $2000\,\textrm{\AA}$.\cite{1983:Beille:SolidStateCommun, 2009:Grigoriev:PhysRevLett} The crystalline lattice constant decreases almost linearly with increasing cobalt content by less than a percent, from $4.48\,\textrm{\AA}$ in FeSi to $4.45\,\textrm{\AA}$ in CoSi.\cite{2004:Manyala:NatureMater} Hydrostatic pressures suppress the magnetic order with critical pressures of several GPa indicating great sensitivity to changes of the lattice constant.\cite{2003:Miura:PhysicaB, 2005:Onose:PhysRevB}

Easy $\langle100\rangle$ axes for the helical propagation vector are associated with the cubic anisotropy, however, less pronounced as for the other cubic chiral magnets especially for large cobalt contents.\cite{1986:Ishimoto:JMagnMagnMater, 1995:Ishimoto:PhysicaB, 2006:Uchida:Science, 2007:Grigoriev:PhysRevB, 2009:Grigoriev:PhysRevLett, 2009:Takeda:JPhysSocJpn} These studies, in fact, were performed prior to the identification of the A phase in Fe$_{1-x}$Co$_{x}$Si as a skyrmion lattice in Ref.~\onlinecite{2010:Munzer:PhysRevB}. In the latter study, for $x = 0.20$ neutron scattering intensity was observed everywhere on a sphere in reciprocal space with a modulus of the helical pitch and unexplained broad maxima around $\langle110\rangle$ akin to the partial order in pure MnSi under hydrostatic pressure.\cite{2004:Pfleiderer:Nature} In addition, it was demonstrated that the magnetic phase diagram is sensitive to the field and temperature history. In particular, field cooling through the skyrmion lattice state may result in a metastable survival down to lowest temperatures. Depending on the field direction, the size of the phase pocket of the skyrmion lattice differs and even two skyrmion lattice domains with different in-plane orientation have been observed.\cite{2010:Adams:JPhysConfSer}

Recently, the metastable skyrmion lattice in Fe$_{1-x}$Co$_{x}$Si was exploited in a study combining magnetic force microscopy and small-angle neutron scattering in order to address the topological unwinding of the skyrmions.\cite{2013:Milde:Science} When the skyrmion lattice transforms into a topologically trivial structure such as the helical state, the emergent magnetic flux discretely vanishes at certain points. These sinks of emergent flux act as a zipper between neighboring skyrmion tubes and have been interpreted as emergent magnetic monopoles. Due to the suppression of thermal fluctuations at low temperatures, the metastable skyrmion lattice in Fe$_{1-x}$Co$_{x}$Si permitted to monitor the intrinsic unwinding mechanisms.

The present study reports comprehensive magnetization, ac susceptibility, and specific heat measurements. We show that Fe$_{1-x}$Co$_{x}$Si, albeit having an electrical resistivity akin to strongly doped semiconductors, may still be discussed as an itinerant magnet. We determine the magnetic phase diagrams over a wide concentration range and for all major crystallographic directions. Despite different transition temperatures and fields, these diagrams are qualitatively very similar across the entire concentration range studied. After zero-field cooling, they resemble that of stoichiometric cubic chiral magnets supporting the helical, conical, and skyrmion lattice phase. At the boundaries between these phases very slow dynamics are observed, comparable to MnSi\cite{2012:Bauer:PhysRevB} or Cu$_{2}$OSeO$_{3}$\cite{2014:Levatic:PhysRevB}. These dynamics are discussed in view of the different response of moments on the scales of the individual atoms and complex topological unwinding processes, respectively. In contrast to stoichiometric compounds such as MnSi, however, the helical state is not recovered in Fe$_{1-x}$Co$_{x}$Si once a large magnetic field has been applied for all field directions. Moreover, when cooled in an applied magnetic field, the skyrmion lattice phase may persist as a metastable state down to the lowest temperatures. Taken together, we show that by choosing a certain cobalt content and applying the appropriate field and temperature history the magnetic properties of Fe$_{1-x}$Co$_{x}$Si may be adjusted over a wide range providing access to interplay of the effects of disorder in various different constellations.

Our paper is organized as follows. After a short summary of the experimental methods in Sec.~\ref{Methods} we present our results in Sec.~\ref{Results}. We start with the compositional phase diagram and an account for the itinerant magnetism of Fe$_{1-x}$Co$_{x}$Si in Sec.~\ref{ItinerantMagnet}. In Sec.~\ref{HelConSkyrmion} we focus on the susceptibility of the helical, conical, and skyrmion lattice state, and the definition of the phase boundaries between these states. As an example, Sec.~\ref{OrientHist} addresses the anisotropy of the magnetic properties and the dependence on the field and temperature history for $x = 0.20$. The resulting magnetic phase diagrams are shown in Sec.~\ref{PhaseDiagram}. Finally, we present measurements of the specific heat in Sec.~\ref{SpecificHeat} before we summarize our findings in Sec.~\ref{Conclusions}.


\section{Experimental Methods}
\label{Methods}

Large single crystals of Fe$_{1-x}$Co$_{x}$Si were prepared in an ultra-high vacuum compatible image furnace.\cite{2011:Neubauer:RevSciInstrum} The polycrystalline feed rods were cast from pure starting elements (4N Fe, 3N5 Co, 6N Si) using a bespoke Hukin crucible with radio-frequency heating inside an all-metal sealed furnace.\cite{2016:Bauer:preprint} The details of the crystal growth process were identical to the growth of Mn$_{1-x}$Fe$_{x}$Si and Mn$_{1-x}$Co$_{x}$Si as described in Ref.~\onlinecite{2010:Bauer:PhysRevB}. Laue X-ray diffraction was used to orient the single crystals. Magnetization and ac susceptibility for magnetic field along $\langle100\rangle$ were measured on cuboids of $1\times1\times6\,\mathrm{mm}^{3}$ with their long edge parallel to $\langle100\rangle$. This way demagnetization effects were minimized due to the small demagnetization factor of $N = 0.07$. The orientation dependence was investigated on cubes of $1\times1\times1\,\mathrm{mm}^{3}$. Heat capacity was measured on quarters of a cylindrical disc with a radius of 3\,mm and a thickness of 1\,mm, where the magnetic field was applied perpendicular to the large surface, i.e., parallel to $\langle110\rangle$.

Magnetization, ac susceptibility, and specific heat were measured in a Quantum Design physical properties measurement system~(PPMS) at temperatures down to 2\,K and in magnetic fields up to 9\,T. The magnetization was determined with an extraction technique. The ac susceptibility was measured at an excitation amplitude of 1\,mT and an excitation frequency of 911\,Hz. The specific heat was measured with a standard heat-pulse method, where typical heat pulses were around 1\% of the temperature at the start of the pulse. All experimental data are shown as a function of applied magnetic field without correction for demagnetization effects, while the phase diagrams inferred from the data are shown as a function of internal field. The demagnetizing factors of the samples were determined by approximating the sample shape as a rectangular prism.\cite{1998:Aharoni:JApplPhys}

The behavior of Fe$_{1-x}$Co$_{x}$Si depends strongly on its temperature and magnetic field history. In turn, it is helpful to distinguish four different scenarios, where all data were recorded between 2\,K and a temperature well above the onset of helimagnetic order. 
\begin{itemize}
	\item Zero-field cooling (zfc): The sample was cooled to 2\,K in zero magnetic field before the desired field value was applied. Data was recorded while increasing the temperature.
	\item Field cooling down (fcd): The sample was cooled to 2\,K in the desired field and data was recorded simultaneously.
	\item Field cooling (fc): The sample was cooled to 2\,K in the desired field and data was recorded while increasing the temperature.
	\item High-field cooling (hfc): The sample was cooled to 2\,K in a field larger than the zero-temperature value of the upper critical field, $H_{c2}$. Subsequently, the desired field was applied and data was recorded while increasing the temperature. 
\end{itemize}
For magnetic field sweeps the sample was initially heated well above the helimagnetic transition temperature before being cooled to the desired temperature in zero field. The data collected during the first increase of the field corresponds to data after zero-field cooling in temperature sweeps. Subsequently, the field was decreased stepwise from $H > H_{c2}$ to $H < -H_{c2}$ and increased stepwise back to $H > H_{c2}$. For the determination of magnetic phase diagrams, data recorded during this loop corresponds to high-field cooling in temperature sweeps.


\section{Experimental results}
\label{Results}

We begin the presentation of our experimental results with the behavior of Fe$_{1-x}$Co$_{x}$Si at zero field and high temperatures as well as at high fields and low temperatures. This provides the setting for the field and temperature dependence of the ac susceptibility observed at low temperatures and small fields, where special emphasis is placed on the determination of the magnetic phase diagrams. The role of the field and temperature history and the dependence on the crystalline orientation are shown in detail for Fe$_{1-x}$Co$_{x}$Si with $x = 0.20$. The results are summarized in magnetic phase diagrams. Finally, we discuss the heat capacity.

\subsection{Itinerant magnetism in Fe$_{1-x}$Co$_{x}$Si}
\label{ItinerantMagnet}

\begin{figure}
\includegraphics[width=1.0\linewidth]{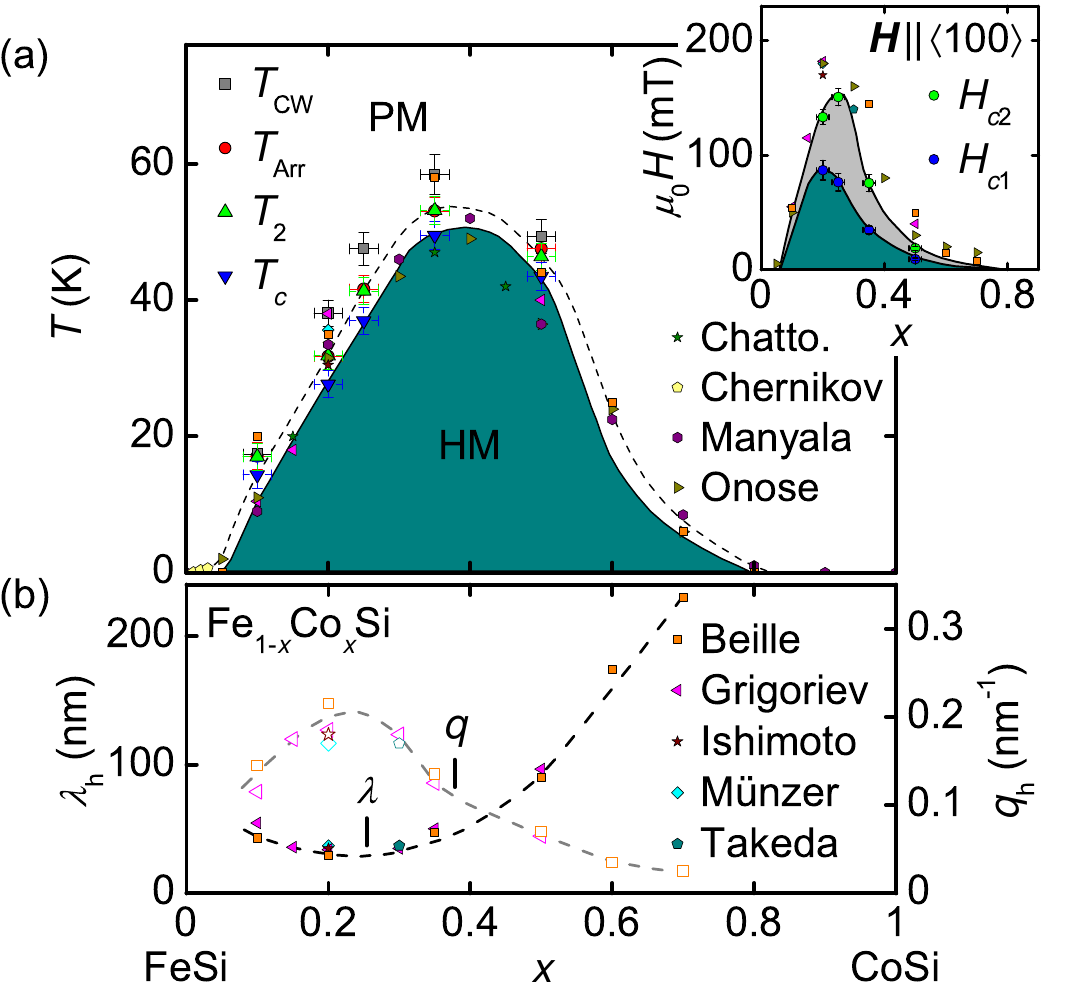}
\caption{(Color online) Helimagnetism in Fe$_{1-x}$Co$_{x}$Si. (a)~Compositional phase diagram. We distinguish a helimagnetic~(HM) and a paramagnetic~(PM) regime. The transition temperatures determined in this study are consistent with earlier reports. The inset shows the low-temperature transition fields as a function of composition. (b)~Helix wavelength, $\lambda_{\mathrm{h}}$, and corresponding wavevector, $q_{\mathrm{h}}$, as a function of cobalt content $x$. The values in panel (a) and (b) are summarized from reports by Beille et al.\cite{1981:Beille:JPhysFMetalPhys, 1983:Beille:SolidStateCommun}, Ishimoto et al.\cite{1995:Ishimoto:PhysicaB}, Chernikov et al.\cite{1997:Chernikov:PhysRevB}, Chattopadhyay et al.\cite{2002:Chattopadhyay:PhysRevB2}, Manyala et al.\cite{2004:Manyala:NatureMater}, Onose et al.\cite{2005:Onose:PhysRevB}, Grigoriev et al.\cite{2007:Grigoriev:PhysRevB, 2007:Grigoriev:PhysRevB2, 2009:Grigoriev:PhysRevLett}, Takeda et al.\cite{2009:Takeda:JPhysSocJpn}, and M\"{u}nzer et al.\cite{2010:Munzer:PhysRevB}.}
\label{figure01}
\end{figure}

We start with the behavior in zero field, where Fe$_{1-x}$Co$_{x}$Si is paramagnetic at high temperatures. At low temperatures, helimagnetic order stabilizes for $0.05 < x < 0.8$. The corresponding helimagnetic transition temperature as a function of the cobalt content $x$ is depicted in Fig.~\ref{figure01}(a). In previous reports\cite{1981:Beille:JPhysFMetalPhys, 1983:Beille:SolidStateCommun, 1995:Ishimoto:PhysicaB, 1997:Chernikov:PhysRevB, 2004:Manyala:NatureMater, 2005:Onose:PhysRevB, 2007:Grigoriev:PhysRevB, 2007:Grigoriev:PhysRevB2, 2009:Grigoriev:PhysRevLett, 2010:Munzer:PhysRevB} this transition temperature was defined in different ways from magnetization, susceptibility, resistivity, and small-angle neutron scattering data. In order to account for this variety, we show the temperatures $T_{c}$ and $T_{2}$ inferred from the ac susceptibility, $T_{\mathrm{CW}}$ inferred from Curie-Weiss plots, and $T_{\mathrm{Arr}}$ inferred from Arrott plots; detailed definitions of these temperatures are given below. Bearing in mind these discrepancies, both the earlier reports and our findings are in very good agreement. 

The inset of Fig.~\ref{figure01}(a) shows the extrapolated zero-temperature values of the transition fields in Fe$_{1-x}$Co$_{x}$Si as a function of the cobalt content. $H_{c1}$ marks the transition between the helical and the conical state and $H_{c2}$ is the transition between the conical and the field-polarized state. Values from earlier reports are typically a few ten millitesla higher than values from the present study. We note that these earlier reports have not been systematic with respect to the crystalline orientation. In Fe$_{1-x}$Co$_{x}$Si, however, as we show below, anisotropies have a relatively large influence on $H_{c2}$ as compared to other cubic chiral magnets. Moreover, demagnetizing effects are sizeable but may not have been taken into account in these reports. In Fig.~\ref{figure01}(b) we finally summarize data available in the literature on the helix wavelength, $\lambda_{\mathrm{h}}$, and the corresponding wavevector, $q_{\mathrm{h}}$. The dependence of $H_{c2}$ and $q_{\mathrm{h}}$ on the cobalt content is qualitatively very similar, with $H_{c2} \propto q_{\mathrm{h}}^{2}$,  as both are determined by the ratio of the coupling constants of Dzyaloshinsky-Moriya and ferromagnetic exchange interaction.

\begin{figure}
\includegraphics[width=1.0\linewidth]{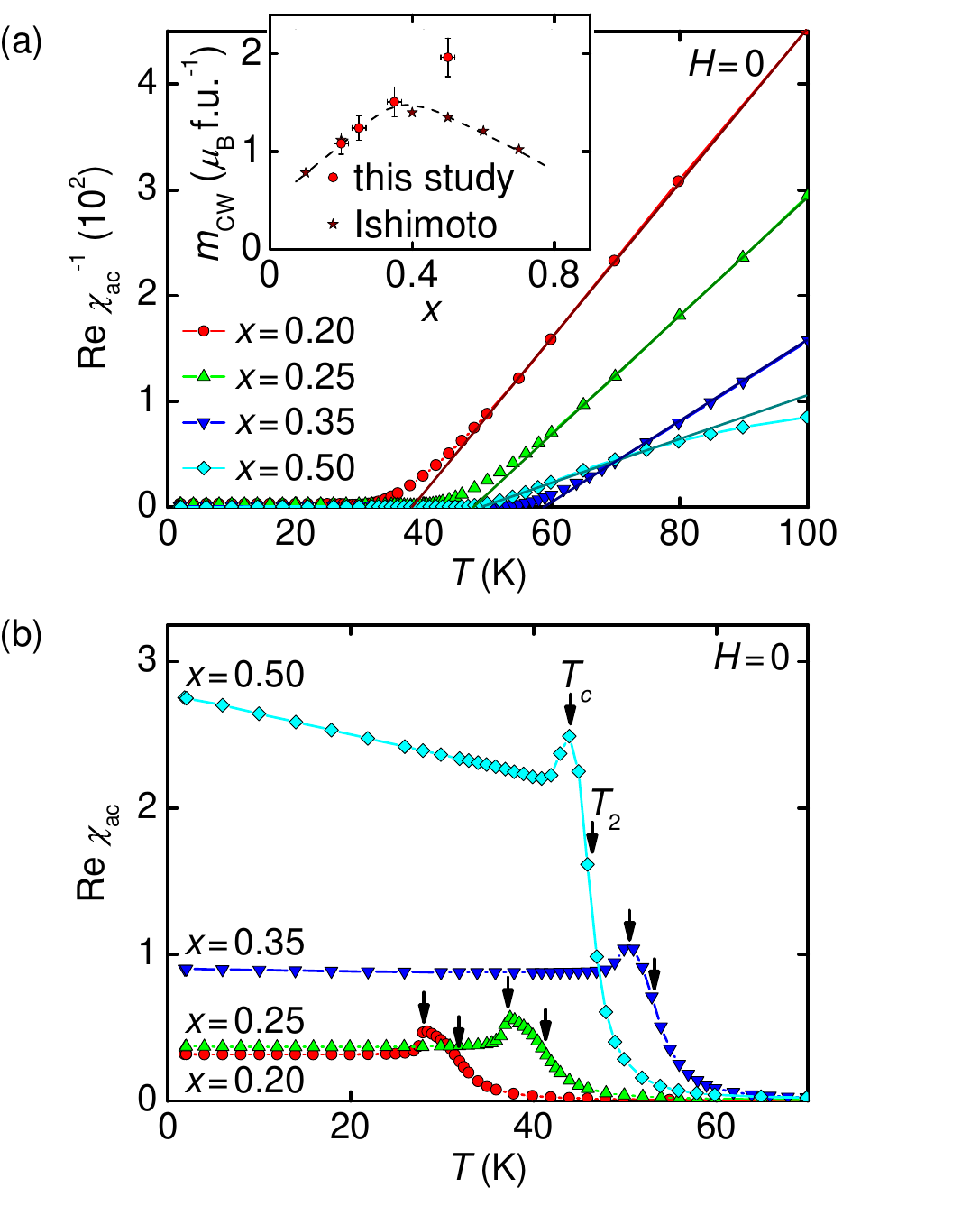}
\caption{(Color online) Magnetic susceptibility of Fe$_{1-x}$Co$_{x}$Si. (a)~Temperature dependence of the inverse ac susceptibility. At high temperatures we observe Curie-Weiss-like behavior. The extrapolated fluctuating moments are shown in the inset. For comparison, we show data from Ishimoto et al.\cite{1992:Ishimoto:JPhysSocJpn}. (b)~Temperature dependence of the real part of the ac susceptibility, Re\,$\chi_{\mathrm{ac}}$. The temperatures $T_{c}$ and $T_{2}$ are defined as kink and point of inflection, respectively.}
\label{figure02}
\end{figure}

The paramagnetic behavior at high temperatures is demonstrated by the inverse ac susceptibility shown in Fig.~\ref{figure02}(a). Well above the helimagnetic transition, we observe a Curie-Weiss-like linear temperature dependence. The Curie-Weiss temperature, $T_{\mathrm{CW}}$, is a few Kelvin higher than the actual helimagnetic transition temperature, as inferred from the ac susceptibility and small-angle neutron scattering. The fluctuating moments, $m_{\mathrm{CW}}$, range between 1\,$\mu_{\mathrm{B}}\,{\mathrm{f.u.}}^{-1}$ and 2\,$\mu_{\mathrm{B}}\,{\mathrm{f.u.}}^{-1}$, in good agreement with the literature\cite{1992:Ishimoto:JPhysSocJpn}. The discrepancy for $x = 0.50$ may be attributed to the different definitions of $m_{\mathrm{CW}}$; while we obtain $m_{\mathrm{CW}}$ from the ac susceptibility in zero field, in Ref.~\onlinecite{1992:Ishimoto:JPhysSocJpn} the moments are determined from measurements of the magnetization in an applied magnetic field of 1\,T.

We now turn to the ac susceptibility at low temperatures shown in Fig.~\ref{figure02}(b). Following a Curie-Weiss-like dependence at high temperatures the susceptibility monotonically increases with decreasing temperature. We observe a point of inflection, marked as $T_{2}$, before the susceptibility drops from its maximum value to a lower plateau. The kink, marked as $T_{c}$, is characteristic of the onset of long-range helimagnetic order. The qualitative shape of the susceptibility of Fe$_{1-x}$Co$_{x}$Si is the same in the concentration range studied. For $x = 0.50$, however, the susceptibility increases linearly for decreasing temperatures within the helical state.

The shape of the susceptibility above $T_{c}$ may be accounted for in the Brazovskii scenario.\cite{1975:Brazovskii:SovPhysJETP, 2013:Janoschek:PhysRevB, 2013:Bauer:PhysRevLett} As the helimagnetic phase transition is approached from high temperatures, ferromagnetic fluctuations gain chiral character thereby occupying a large volume in reciprocal space. These fluctuations may efficiently interact with each other and suppress the second-order phase transition expected in mean-field theory. With the suppression of the helimagnetic transition a fluctuation-disordered regime forms below the point of inflection at $T_{2}$ down to the fluctuation-induced first-order phase transition at $T_{c}$.

\begin{figure}
\includegraphics[width=1.0\linewidth]{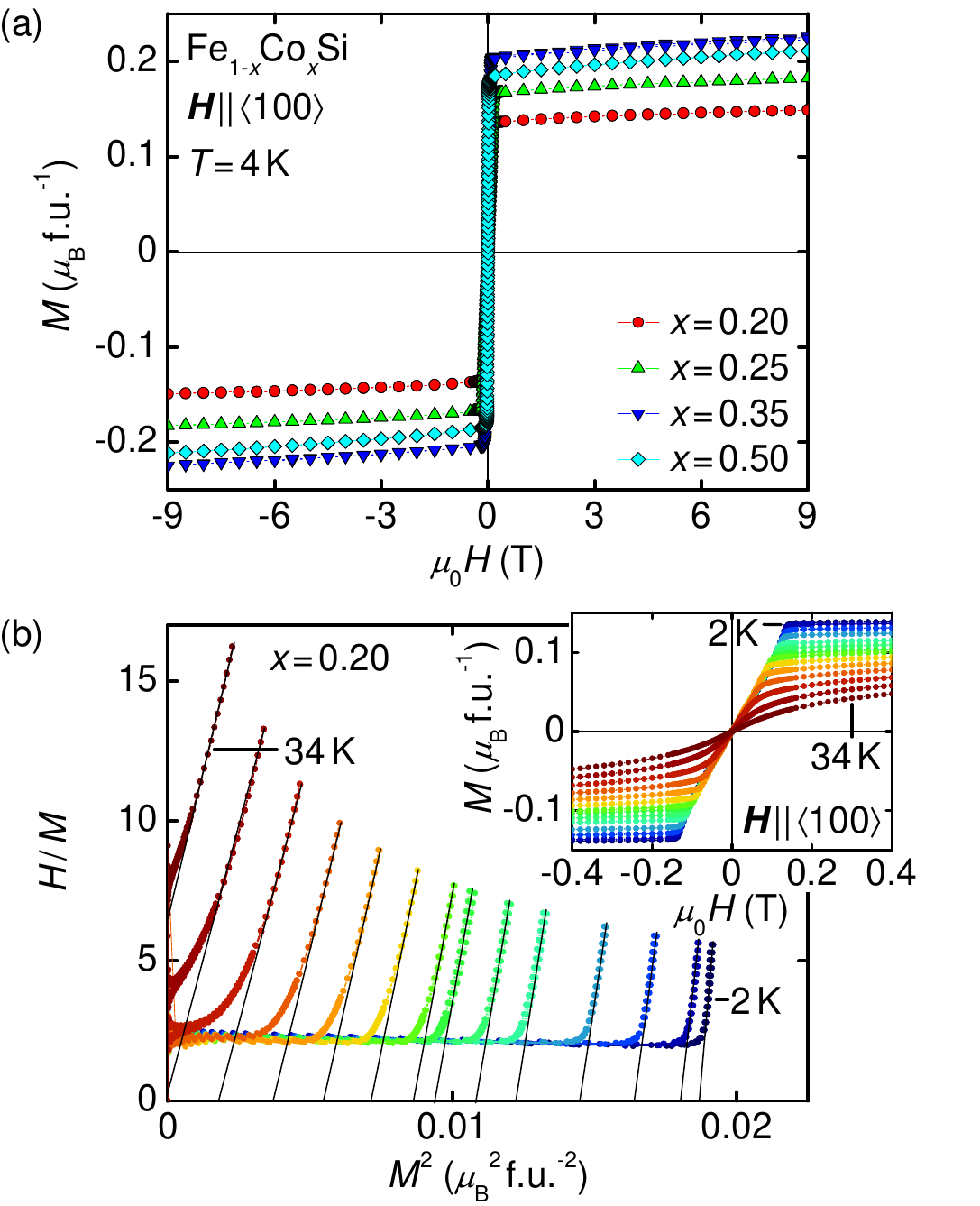}
\caption{(Color online) Field dependence of the magnetization of Fe$_{1-x}$Co$_{x}$Si at high fields and low temperatures. (a)~Isothermal magnetization at 4\,K as a function of field up to 9\,T for different cobalt contents $x$. (b)~Arrott plot, i.e., magnetic field divided by magnetization versus the square of the magnetization, for $x = 0.20$ and several temperatures. Solid lines represent linear fits to the high-field data. The inset displays the corresponding isothermal magnetization.}
\label{figure03}
\end{figure}

In the following, we describe the magnetization as a function of field, see Fig.~\ref{figure03}. The steep increase of the isothermal magnetization, $M$, around zero field is associated with the helimagnetism in Fe$_{1-x}$Co$_{x}$Si and will be discussed in detail in terms of the susceptibility in Sec.~\ref{HelConSkyrmion}. Here, we focus on the behavior at high fields. For all concentrations studied, the magnetization is of the order of 0.2\,$\mu_{\mathrm{B}}\,{\mathrm{f.u.}}^{-1}$ and unsaturated up to 9\,T. The lack of saturation of the spin-polarized state up to the largest fields measured is a characteristic of itinerant magnetism and consistent with MnSi, where an unsaturated and nonlinear magnetization has been reported up to 33\,T.\cite{1982:Sakakibara:JPhysSocJpn}

For an itinerant ferromagnet, the equation of state on the mean-field level is given by $B = aM + bM^{3}$.\cite{1985:Lonzarich:JPhysCSolidState} In order to infer the spontaneous ordered moment for a given temperature, $m_{s}$, the isothermal magnetization is plotted as $H/M = B/\mu_{0}M$ versus $M^{2}$. Subsequently, the high-field behavior, where helimagnetism is suppressed, is linearly extrapolated to $H = 0$ providing $m_{s}$. This so-called (inverted) Arrott plot is shown for $x = 0.20$ in Fig.~\ref{figure03}(b) for several different temperatures, where the corresponding magnetization data are depicted in the inset. For weak itinerant ferromagnets the spontaneous moment is expected to vary with temperature as $m_{s}^{2} = m_{s,0}^{2}\left(1-(T/T_{\mathrm{Arr}})^{\alpha}\right)$ with the zero-temperature moment, $m_{s,0}$, and $\alpha = 2$. The behavior of the doped helimagnets Mn$_{1-x}$Fe$_{x}$Si and Mn$_{1-x}$Co$_{x}$Si is in excellent agreement with this prediction, consistent with itinerant spin fluctuations.\cite{2010:Bauer:PhysRevB, 1979:Moriya:JMagnMagnMater, 1985:Lonzarich:JPhysCSolidState, 1986:Lonzarich:JMagnMagnMater, 1991:Moriya:JMagnMagnMater} 

\begin{figure}
\includegraphics[width=1.0\linewidth]{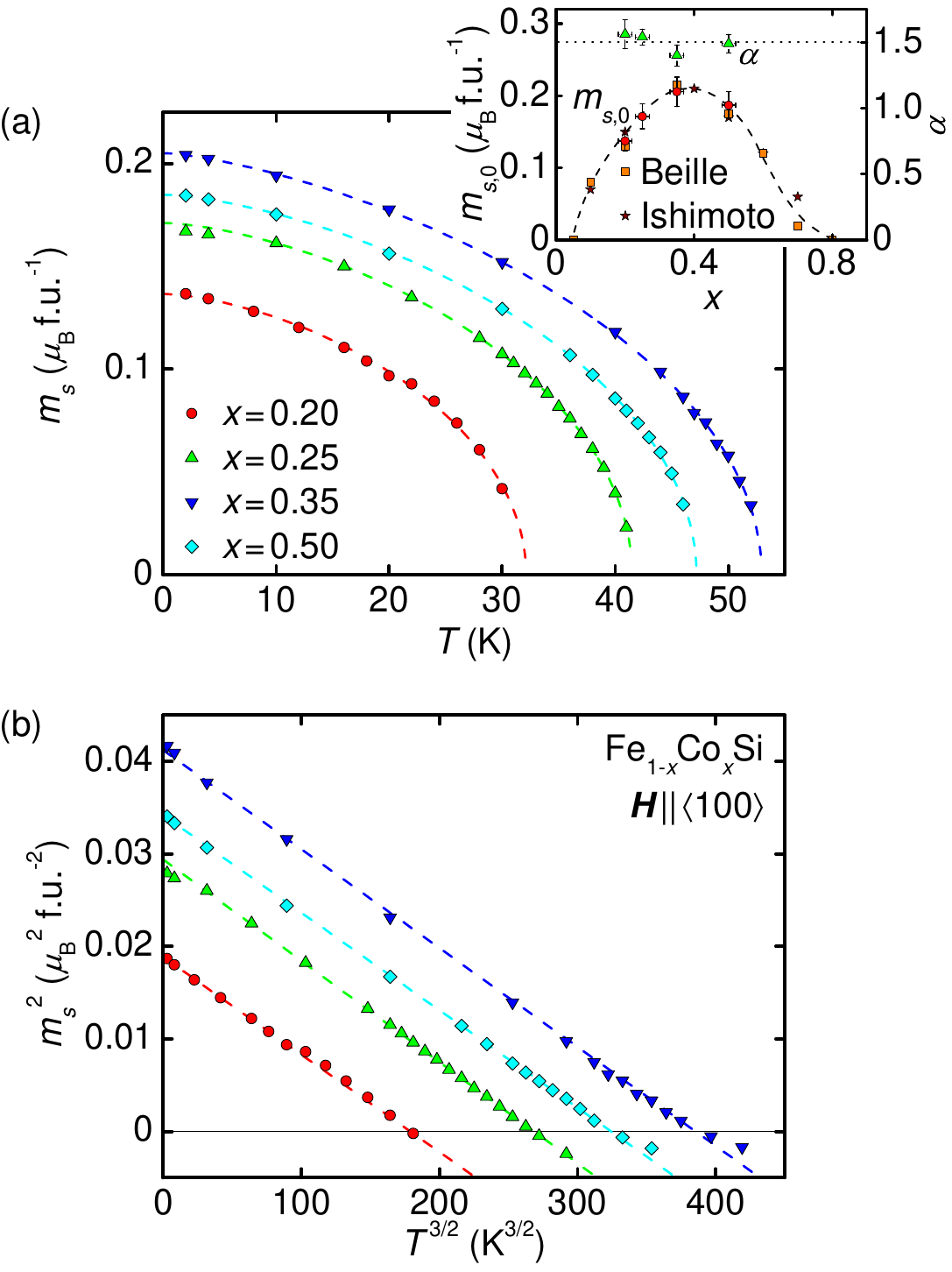}
\caption{(Color online) Key properties of Fe$_{1-x}$Co$_{x}$Si at high fields and low temperatures. (a)~Temperature dependence of the spontaneous moment, $m_{s}$, as extrapolated from Arrott plots. The dashed lines are fits following $m_{s}^{2} = m_{s,0}^{2}\left(1-(T/T_{\mathrm{Arr}})^{\alpha}\right)$. The inset shows the concentration dependence of the zero-temperature spontaneous moment, $m_{s,0}$. For comparison, we show data from Beille et al.\cite{1983:Beille:SolidStateCommun} and Ishimoto et al.\cite{1992:Ishimoto:JPhysSocJpn}. The exponent $\alpha = 3/2$ differs from typical itinerant ferromagnets where one expects $\alpha = 2$. (b)~Square of the spontaneous moment, $m_{s}^{2}$, as a function of $T^{3/2}$.}
\label{figure04}
\end{figure}

For Fe$_{1-x}$Co$_{x}$Si, we show the spontaneous moment as a function of temperature in Fig.~\ref{figure04}(a). The dashed lines represent fits according to the abovementioned formula and with the free parameters $T_{c}$, $m_{s,0}$, and $\alpha$, describing the experimental data very well. The characteristic temperatures $T_{\mathrm{Arr}}$ from these fits are shown in Fig.~\ref{figure01}(a). Their values coincide with the temperature $T_{2}$ inferred from the point of inflection of the ac susceptibility as expected within the Brazovskii scenario.\cite{2013:Janoschek:PhysRevB} The spontaneous zero-temperature ordered moments, $m_{s,0}$, are depicted in the inset of Fig.~\ref{figure04}(a). They are roughly one order of magnitude smaller than the fluctuating Curie-Weiss moments, as characteristic for weak itinerant ferromagnetism. Both $T_{\mathrm{Arr}}$ and $m_{s,0}$ reach their maximum for a cobalt content of $x \approx 0.35$. The inset of Fig.~\ref{figure02}(a) also depicts the exponent $\alpha$. We observe $\alpha = 3/2$ for all samples studied which is highlighted in Fig.~\ref{figure04}(b), where we show the square of the spontaneous moment, $m_{s}^{2}$, exhibiting a linear relationship as a function of $T^{3/2}$. The value $\alpha = 3/2$ indicates that the temperature dependence of the extrapolated value of $m_{s}$ is governed by the thermal excitation of spin-waves at temperatures much larger than the spin-wave gap, $k_{\mathrm{B}}T \gg \Delta$.\cite{1985:Lonzarich:JPhysCSolidState} This finding is in contrast to single-electron excitations that yield the exponent $\alpha = 2$ which was previously observed in polycrystals\cite{2002:Chattopadhyay:PhysRevB2}.

\subsection{Magnetically modulated phases}
\label{HelConSkyrmion}

\begin{figure}
\includegraphics[width=1.0\linewidth]{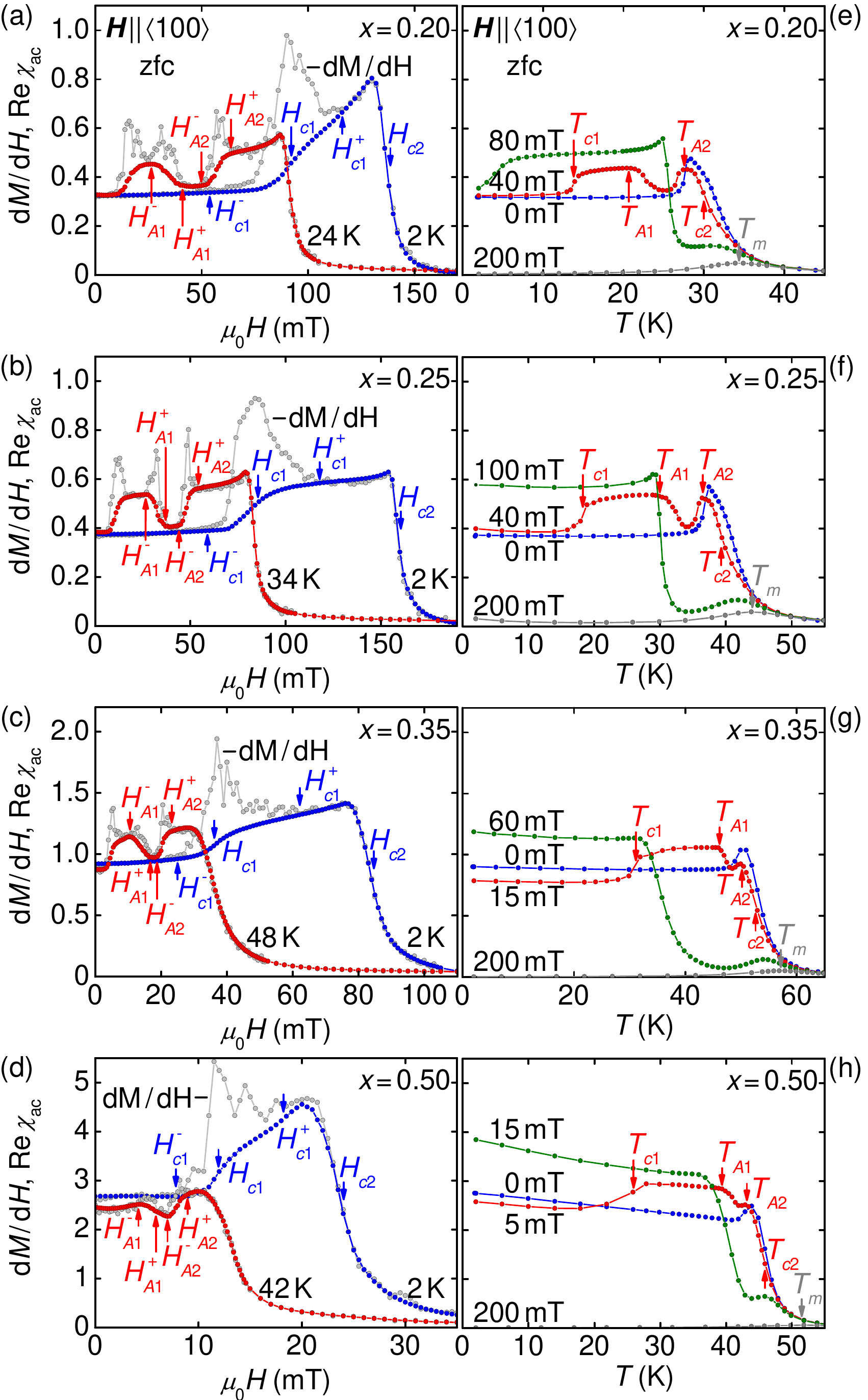}
\caption{(Color online) Typical field and temperature dependence of the susceptibility of Fe$_{1-x}$Co$_{x}$Si for various concentrations after zero-field cooling. The magnetic field was parallel $\langle100\rangle$. \mbox{(a)--(d)}~Field dependence at 2\,K (blue) and at a temperature a few Kelvin below $T_{c}$ (red). The derivative of the dc magnetization, $\mathrm{d}M/\mathrm{d}H$, is shown in gray. \mbox{(e)--(h)}~Temperature dependence for zero field (blue) as well as for typical field values in the skyrmion lattice state (red), the conical state (green), and the field-polarized regime (gray). See text for definitions of the transition fields and temperatures.}
\label{figure05}
\end{figure}

In the following, we present the behavior of Fe$_{1-x}$Co$_{x}$Si at low temperatures and in small magnetic fields. The various magnetically modulated phases in this part of the phase diagram may be discriminated best by means of the susceptibility. Figure~\ref{figure05}(a) shows typical field dependencies of the susceptibility calculated from the magnetization, $\mathrm{d}M/\mathrm{d}H$, and the real part of the ac susceptibility, Re\,$\chi_{\mathrm{ac}}$, for Fe$_{1-x}$Co$_{x}$Si with $x = 0.20$. Only data for the lowest temperature measured, 2\,K, and a temperature a few Kelvin below the helimagnetic transition temperature, $T_{c}$, are shown for clarity. The magnetic field was applied along the crystalline $\langle100\rangle$ direction after zero-field cooling. The nomenclature of the transition fields and temperatures will be presented in the following along with the account of the results, consistent with the definitions given in Ref.~\onlinecite{2012:Bauer:PhysRevB}.

The description starts with the susceptibility calculated from the magnetization, $\mathrm{d}M/\mathrm{d}H$, at 2\,K and zero field. Here, Fe$_{1-x}$Co$_{x}$Si is in the helical state. For increasing field, $\mathrm{d}M/\mathrm{d}H$ is nearly unchanged for $H < H_{c1}^{-}$. At $H_{c1}^{-}$ a pronounced peak emerges that reaches its maximum at $H_{c1}$ and vanishes at $H_{c1}^{+}$. For $H > H_{c1}^{+}$ Fe$_{1-x}$Co$_{x}$Si is in the conical state. Other than for pure or doped MnSi, $\mathrm{d}M/\mathrm{d}H$ is not independent of the field in this part of the phase diagram. If the field is increased further, the system finally enters the field-polarized state for $H > H_{c2}$, where $H_{c2}$ is defined as point of inflection. The real part of the ac susceptibility is consistent with $\mathrm{d}M/\mathrm{d}H$, except that it does not track the peak at the helical-to-conical phase transition around $H_{c1}$. The latter is also observed in MnSi providing evidence of very slow dynamics at the phase boundary.

In general, an account for the magnetic susceptibility of the cubic chiral magnets requires consideration of two contributions, namely the response of magnetic moments on atomic length scales and the entire helix, respectively. The former contribution is fast and may be probed by both the ac susceptibility and the susceptibility calculated from the magnetization. In contrast, the response of the propagation vector of the helix is slow and only leads to sizable contributions around the helical-to-conical phase transition where the helix reorients as a function of field. The associated time scale depends on the temperature and is typically of the order of milliseconds. A detailed discussion of these processes will be presented elsewhere.\cite{2016:Bauer:preprint2} 

For temperatures a few Kelvin below the helimagnetic transition the description given above remains valid. Additionally, two peaks bordering a plateau of reduced susceptibility emerge in $\mathrm{d}M/\mathrm{d}H$ at intermediate fields. This signature is the characteristic of the skyrmion lattice phase. The peaks are not tracked by the ac susceptibility measured at an excitation frequency of 911\,Hz indicating slightly smeared first-order transitions with very slow dynamics and finite dissipation, consistent with the behavior seen in MnSi.\cite{2012:Bauer:PhysRevB, 2013:Bauer:PhysRevLett} The slow processes at this phase boundary are associated with the nucleation process of skyrmions within the topologically trivial conical phase and vice versa.\cite{2013:Milde:Science, 2015:Rajeswari:PNatlAcadSciUSA} For $x = 0.25$, $x = 0.35$, and $x = 0.50$, as shown in Figs.~\ref{figure05}(b) through \ref{figure05}(d), the susceptibility is highly reminiscent of $x = 0.20$. The critical field and temperature values, however, change considerably.

The picture drawn from the field dependence of the susceptibility is corroborated by the temperature dependence after zero-field cooling. We start our description with $x = 0.20$ depicted in Fig.~\ref{figure05}(e). In zero magnetic field below $T_{c}$ Fe$_{1-x}$Co$_{x}$Si is in the helical state. Above $T_{c}$ the system is paramagnetic, where the point of inflection at $T_{2}$ marks the crossover from the fluctuation-disordered to the mean-field disordered regime as described by the Brazovskii scenario.\cite{2013:Janoschek:PhysRevB}

Under magnetic field, an additional point of inflection at $T_{c1}$ is attributed to the transition from the helical state at low temperatures to the conical state at higher temperatures. This transition is observed in Fe$_{1-x}$Co$_{x}$Si as the critical field of the helical-to-conical transition, $H_{c1}$, shows a pronounced temperature dependence, other than in pure MnSi but similar to Mn$_{1-x}$Fe$_{x}$Si or Mn$_{1-x}$Co$_{x}$Si.\cite{2010:Bauer:PhysRevB} Note that the zero-field values of $T_{c1}$ and $T_{c2}$ are also referred to as $T_{1}$ and $T_{2}$, respectively. As $T_{c}$ indicates the temperature at which long-range magnetic order sets in; in zero field $T_{c} = T_{1}$.

For intermediate fields an additional minimum appears at high temperatures within the conical plateau. This minimum coincides with the minimum observed in the field dependence of the susceptibility and is characteristic of the skyrmion lattice state. The transition temperatures, $T_{A1}$ and $T_{A2}$, are defined as the beginning and the end of the deviation of the susceptibility in the conical state, respectively.

For higher fields the plateau in the conical state remains unchanged and a shallow maximum emerges in the susceptibility at a temperature $T_{m} > T_{c2}$. This maximum marks the crossover between the field-polarized state at low temperatures and the paramagnetic state at high temperatures. For $H > H_{c2}$ only the maximum survives and shifts to higher temperatures with increasing magnetic fields. The behavior of $x = 0.25$, $x = 0.35$, and $x = 0.50$, as shown in Figs.~\ref{figure05}(f) through \ref{figure05}(h), is again highly reminiscent of $x = 0.20$.

\subsection{Orientation and history dependence}
\label{OrientHist}

\begin{figure*}
\includegraphics[width=1.0\linewidth]{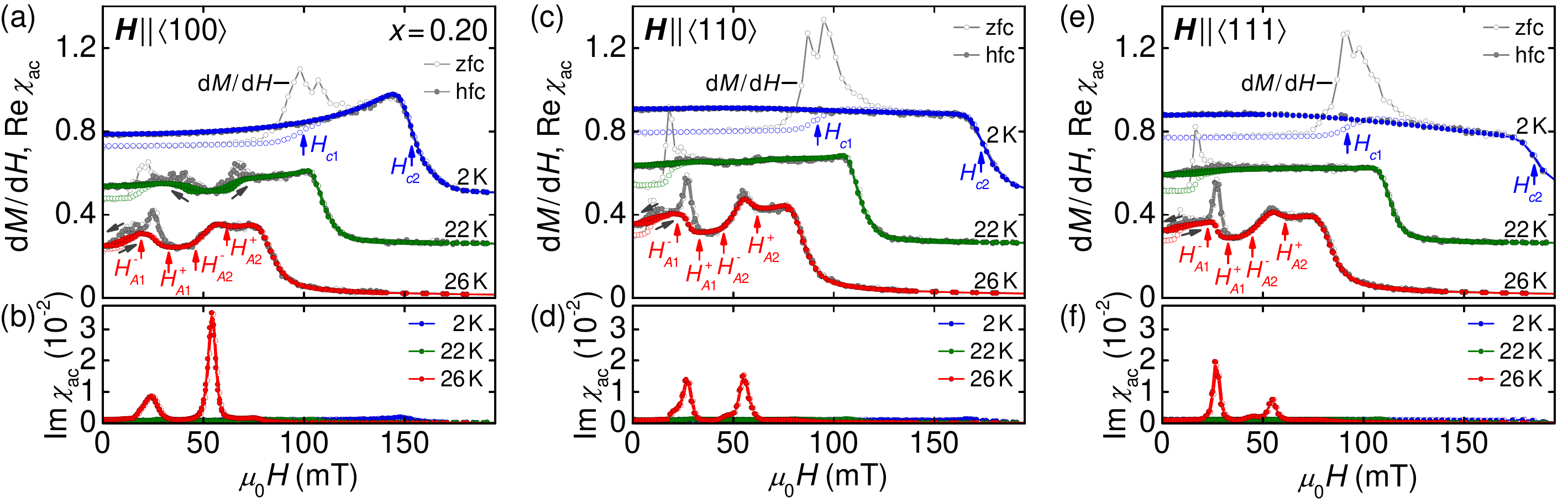}
\caption{(Color online) Typical field dependence of the susceptibility of Fe$_{1-x}$Co$_{x}$Si with $x = 0.20$ for different field directions. Data are shown after zero-field cooling (zfc, open symbols) and high-field cooling (hfc, solid symbols) for 2\,K (blue) and two temperature just below $T_{c}$ (green, red). (a)~Real part of the ac susceptibility, Re\,$\chi_{\mathrm{ac}}$, for field along $\langle100\rangle$. The derivative of the dc magnetization, $\mathrm{d}M/\mathrm{d}H$, is shown in gray. Data are offset by 0.25 for clarity. (b)~Imaginary part of the ac susceptibility, Im\,$\chi_{\mathrm{ac}}$, for field along $\langle100\rangle$. \mbox{(c)--(f)}~Susceptibility for field along $\langle110\rangle$ and $\langle111\rangle$. The overall behavior is very similar for all major crystallographic directions.}
\label{figure06}
\end{figure*}

In the following we consider the influence of the crystalline orientation as well as of the field and temperature history on the magnetic phase diagram of Fe$_{1-x}$Co$_{x}$Si. A detailed account is presented for $x = 0.20$ only as similar behavior is observed for all concentrations studied (not shown). Our description starts with the susceptibility as a function of field and continues with data as a function of temperature. 

Figure~\ref{figure06} shows the field dependence of the susceptibility for field along $\langle100\rangle$, $\langle110\rangle$, and $\langle111\rangle$. For clarity, data are shown for the lowest temperature measured, 2\,K, as well as for two temperatures around the lower and upper temperature boundary of the skyrmion lattice phase, at 22\,K and 26\,K, respectively. Open symbols indicate the behavior after zero-field cooling as already addressed in Fig.~\ref{figure05}. Solid symbols show the behavior following the application of a large positive or negative field, $\left|H\right| > H_{c2}$, corresponding to the situation after high-field cooling in temperature sweeps.

In general, the magnetic properties of Fe$_{1-x}$Co$_{x}$Si are rather isotropic, whereas distinct discrepancies exist between the different field histories. At first, we focus on the helical state as studied by the real part of the ac susceptibility, Re\,$\chi_{\mathrm{ac}}$, and the susceptibility calculated from the magnetization, $\mathrm{d}M/\mathrm{d}H$, see Figs.~\ref{figure06}(a), \ref{figure06}(c), and \ref{figure06}(e). After zero-field cooling, the signature of the helical-to-conical transition and in particular the value of the transition field, $H_{c1}$, is essentially independent of orientation. Following the application of a magnetic field, however, at low temperatures there is no sign of the helical-to-conical transition for all field directions, in agreement with previous small-angle neutron scattering studies\cite{1986:Ishimoto:JMagnMagnMater, 2007:Grigoriev:PhysRevB, 2009:Takeda:JPhysSocJpn, 2010:Munzer:PhysRevB}. Both findings contrast stoichiometric compounds such as MnSi and Cu$_{2}$OSeO$_{3}$ but are consistent with doped systems such as Mn$_{1-x}$Fe$_{x}$Si. They suggest overall weak cubic anisotropies in combination with pronounced local pinning of helices due to structural disorder, where the latter is expected to play an important role in Fe$_{1-x}$Co$_{x}$Si in general.\cite{2006:Punkkinen:PhysRevB, 2007:Racu:PhysRevB} Just below $T_{c}$, however, a minimum emerges around zero field that resembles the minimum observed after zero-field cooling. In turn, we associate the latter signature with the re-population of a multi-domain helical state due to an increase of thermal fluctuations as also observed in other cubic chiral magnets.\cite{2010:Bauer:PhysRevB} The minimum is most pronounced for field along $\langle100\rangle$ and shows weak hysteresis between sweeps for increasing and decreasing field as indicated by the gray arrows.

While $H_{c1}$ is essentially isotropic, the transition between the conical and the field-polarized state at $H_{c2}$ exhibits pronounced anisotropy. At low temperatures, $H_{c2}$ increases by about 20\% from field parallel $\langle100\rangle$ to field parallel $\langle111\rangle$. This anisotropy is much larger than for pure or doped MnSi. Still, we observe no hysteresis at $H_{c2}$ within the resolution of our study. The shape of the susceptibility in the conical state slightly changes for different field directions.

At 26\,K a plateau of reduced susceptibility at intermediate fields is characteristic of the formation of the skyrmion lattice state for all field directions and cooling histories studied. At 22\,K, we observe no corresponding signatures for field along $\langle110\rangle$ and $\langle111\rangle$, whereas the skyrmion lattice is still present for field along $\langle100\rangle$. In the latter case, the sweeps for increasing and decreasing field slightly deviate from each other where the sweep for increasing field tracks data recorded after zero-field cooling. Taken together, the temperature range of the skyrmion lattice is largest for field parallel $\langle100\rangle$ and, as we will show in the temperature sweeps below, smallest for field parallel $\langle111\rangle$.

\begin{figure*}
\includegraphics[width=1.0\linewidth]{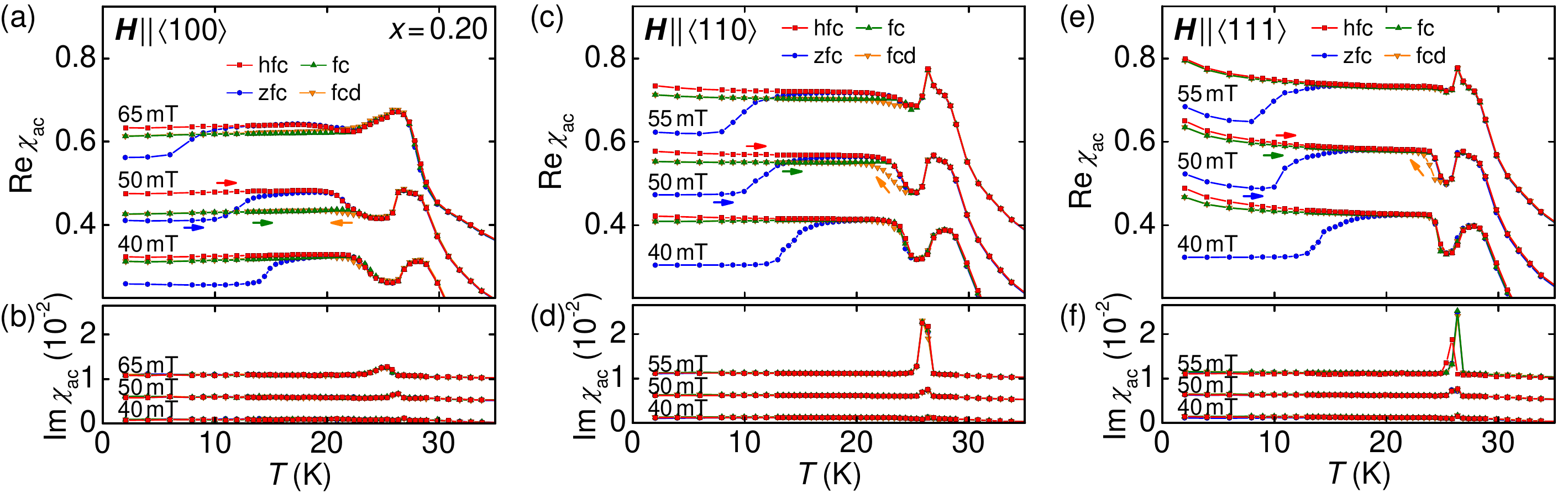}
\caption{(Color online) Typical temperature dependence of the susceptibility of Fe$_{1-x}$Co$_{x}$Si with $x = 0.20$ for different field directions. Data are shown for three applied field values and increasing temperatures after zero-field cooling (zfc, blue), field cooling (fc, green), and high-field cooling (hfc, red) as well as for decreasing temperatures during field cooling (fcd, orange). (a)~Real part of the ac susceptibility, Re\,$\chi_{\mathrm{ac}}$, for field along $\langle100\rangle$. Data are offset by 0.15 for clarity. (b)~Imaginary part of the ac susceptibility, Im\,$\chi_{\mathrm{ac}}$, for field along $\langle100\rangle$. Data are offset by 0.005 for clarity. \mbox{(c)--(f)}~Susceptibility for field along $\langle110\rangle$ and $\langle111\rangle$. The overall behavior is very similar for all major crystallographic directions.}
\label{figure07}
\end{figure*}

The imaginary part of the ac susceptibility, Im\,$\chi_{\mathrm{ac}}$, is shown in Figs.~\ref{figure06}(b), \ref{figure06}(d), and \ref{figure06}(f) and is very similar for all field directions. It is small except for pronounced peaks at the transitions between the skyrmion lattice and the conical state. The finite dissipation may be attributed to a regime of phase coexistence and is also observed in other cubic chiral magnets.\cite{2012:Bauer:PhysRevB} Within the resolution of our study there are no differences between zero-field and high-field cooling.

Further insights into the history dependence of Fe$_{1-x}$Co$_{x}$Si may be inferred from the temperature dependence of the susceptibility as depicted in Fig.~\ref{figure07} for field along $\langle100\rangle$, $\langle110\rangle$, and $\langle111\rangle$. Here, we focus on field values in the skyrmion lattice phase and distinguish four different cooling histories, notably (i)~zero-field cooling (zfc, blue), (ii)~field cooling down (fcd, orange), (iii)~field cooling (fc, green), and (iv)~high-field cooling (hfc, red).

We start with the real part of the ac susceptibility, Re\,$\chi_{\mathrm{ac}}$, shown in Figs.~\ref{figure07}(a), \ref{figure07}(c), and \ref{figure07}(e). As already discussed before, the helical state at low temperatures only forms after zero-field cooling. The signature of the skyrmion lattice state is a minimum at the high-temperature side of the conical plateau. In this pocket just below the onset of long-range order a stable skyrmion lattice forms irrespective of the field and temperature history. When field cooling the minimum may be extended to lower temperatures. Around its low field boundary the skyrmion lattice phase may be enlarged by a few Kelvin, while at higher fields the metastable regime may persist down to the lowest temperatures studied. Both the temperature width of the pocket of the skyrmion lattice phase and the magnitude of its metastable extension are largest for field along $\langle100\rangle$ and smallest for $\langle111\rangle$. The observed phase boundaries, both stable and metastable, are consistent with small-angle neutron scattering data.\cite{2010:Munzer:PhysRevB} After high-field cooling, the helical state is suppressed and the skyrmion lattice state is only present in the stable phase pocket. We note that as a result, for instance at a field of $\mu_{0}H = 50$\,mT along $\langle100\rangle$, either the helical state (after zero-field cooling), the skyrmion lattice state (after field cooling), or the conical state (after high-field cooling) may be observed at low temperatures. 

Recently, also the skyrmion lattice in MnSi was metastably cooled down to low temperatures at ambient pressure by exploiting electric heating and subsequent rapid cooling\cite{2016:Oike:NaturePhys} as well as under hydrostatic pressure.\cite{2013:Ritz:PhysRevB, 2013:Ritz:Nature} Presumably due to the disorder present in the system, in Fe$_{1-x}$Co$_{x}$Si the time scales associated with the unwinding of the skyrmion lattice\cite{2013:Milde:Science} are distinctly larger than in MnSi. Thus, typical cooling rates of the order of 1\,K\,min$^{-1}$ are already fast enough to prevent the decay of the skyrmion lattice at its phase boundary to the conical state. At low temperatures, fluctuations are finally no longer sufficient to initiate the unwinding. 

The imaginary part of the ac susceptibility, Im\,$\chi_{\mathrm{ac}}$, depicted in Figs.~\ref{figure07}(b), \ref{figure07}(d), and \ref{figure07}(f), is independent of the field and temperature history at the resolution of the present study. Consistent with the data as a function of field, the imaginary part of the ac susceptibility shows peaks around the phase boundary of the skyrmion lattice state, while it is negligible everywhere else.

\subsection{Magnetic phase diagrams}
\label{PhaseDiagram}

\begin{figure}
\includegraphics[width=1.0\linewidth]{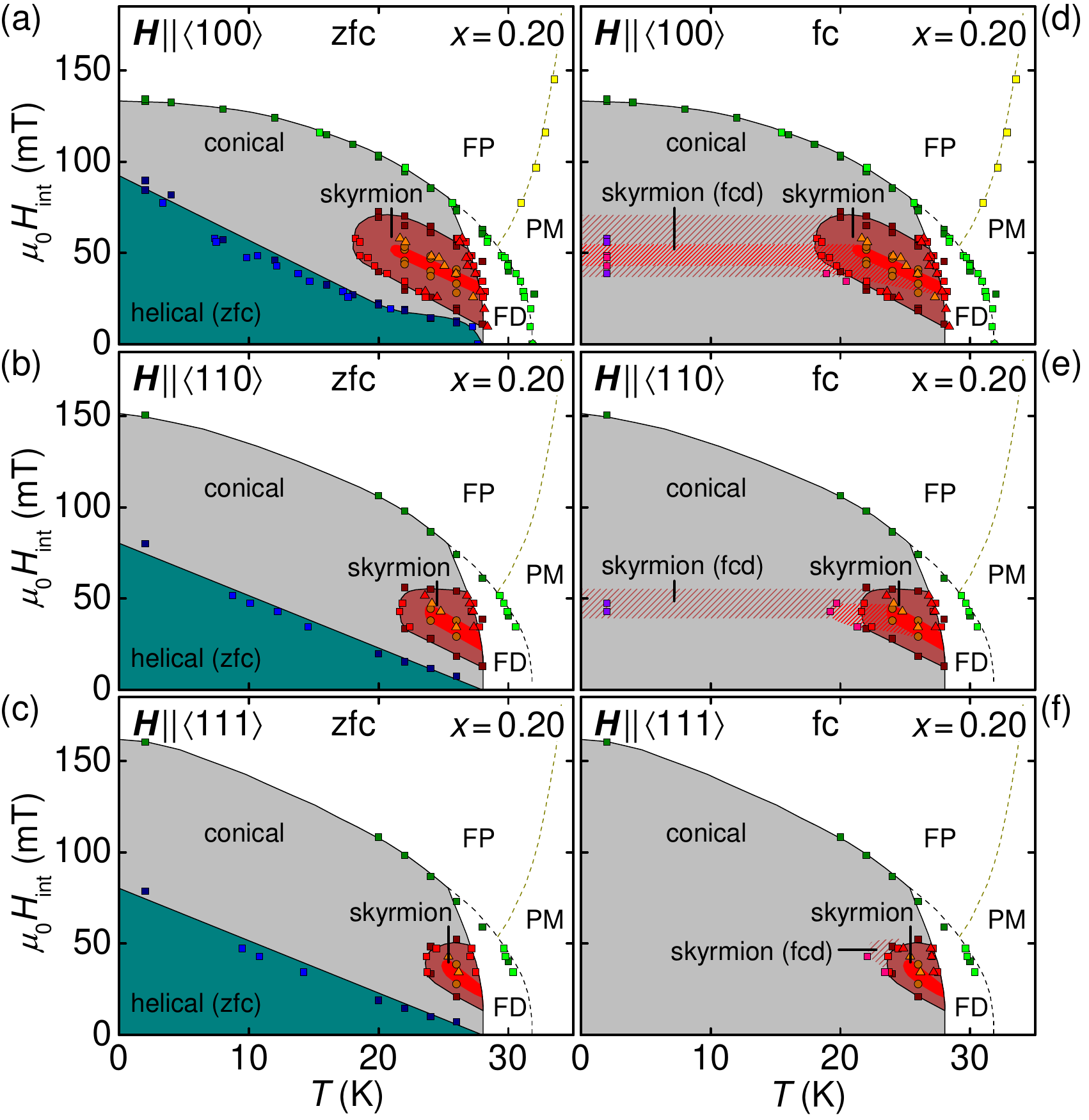}
\caption{(Color online) Magnetic phase diagram of Fe$_{1-x}$Co$_{x}$Si with $x = 0.20$ for field along $\langle100\rangle$, $\langle110\rangle$, and $\langle111\rangle$ after zero-field cooling (left) and field cooling (right). Data are shown as a function of internal field, i.e., after correcting for demagnetization effects. We distinguish six regimes: helical, conical, skyrmion lattice, paramagnetic~(PM), field-polarized~(FP), and fluctuation-disordered~(FD). The overall behavior is very similar for the three directions. The helical state is only observed after zero-field cooling. Under field cooling the skyrmion lattice may be extended metastably down to low temperatures.}
\label{figure08}
\end{figure}

In the following, we present the magnetic phase diagrams as inferred from the susceptibility data. Figure~\ref{figure08} summarizes our findings for Fe$_{1-x}$Co$_{x}$Si with $x = 0.20$ under zero-field cooling (left column) and field cooling (right column) for magnetic field along the major crystallographic axes.

After zero-field cooling, the magnetic phase diagram resembles that of the archetypical cubic chiral magnet MnSi. At high temperatures and low fields, Fe$_{1-x}$Co$_{x}$Si is paramagnetic~(PM), at low temperatures and high fields it is field-polarized~(FP). The two regimes are separated by a crossover at $T_{m}$ inferred from a shallow maximum in the ac susceptibility. At the transition from the paramagnetic to the long-range modulated states a fluctuation-disordered regime~(FD) is observed that is dominated by strongly interacting chiral fluctuations.\cite{2013:Janoschek:PhysRevB}

Below the helimagnetic ordering temperature, $T_{c}$, we observe a helical state at low fields, a conical state at higher fields, and a skyrmion lattice state at intermediate fields just below $T_{c}$. The transition regimes between these phases are characterized by very slow dynamics. At the helical-to-conical transition, where we only show $H_{c1}$ and omit $H_{c1}^{\pm}$ for clarity, this effect arises from the slow reorientation of the helical propagation vector. Around the skyrmion lattice we attribute the slow response to the nucleation and topological unwinding processes of the skyrmions resulting in a regime of phase coexistence with finite dissipation, as typically observed at first-order phase boundaries.

A stable skyrmion lattice independent of the cooling history is realized in a single phase pocket just below $T_{c}$. The temperature extent of this pocket is largest for field along $\langle100\rangle$ and smallest for field along $\langle111\rangle$. The upper critical field of the skyrmion lattice decreases with increasing temperature giving rise to a reentrant conical state. For all crystallographic directions the temperature extent of the skyrmion lattice state, especially in relation to $T_{c}$, is large compared to stoichiometric compounds exhibiting a skyrmion lattice.

Other than in MnSi, the critical field of the helical-to-conical transition, $H_{c1}$, increases in Fe$_{1-x}$Co$_{x}$Si with decreasing temperature and is essentially independent of the crystallographic orientation. Instead, the transition between the conical and the field-polarized state at $H_{c2}$ shows a comparably strong anisotropy of about 20\% with the smallest value for field along $\langle100\rangle$. Still, the anisotropy of the skyrmion lattice phase implies an easy $\langle111\rangle$ axes for the helical pitch in Fe$_{1-x}$Co$_{x}$Si with $x = 0.20$, although $H_{c1}$ shows no direction dependence. In contrast, in small-angle neutron scattering intensity maxima along the $\langle110\rangle$ axes were detected after zero-field cooling.\cite{2010:Munzer:PhysRevB} In combination with the pronounced anisotropy of $H_{c2}$, these observations suggest that nominally sub-leading crystalline anisotropies have a relatively strong effect on the magnetic ordering.

Under field cooling, two distinct differences are observed with respect to the behavior after zero-field cooling. First, the helical state is not recovered once a field larger than $H_{c1}$ has been applied for all crystallographic directions. Similar observations in Mn$_{1-x}$Fe$_{x}$Si and Mn$_{1-x}$Co$_{x}$Si\cite{2010:Bauer:PhysRevB} suggest that the disorder introduced by the doping has a strong influence on the magnetic anisotropy of the helices and the formation of helical domain walls. Second, under field cooling the skyrmion lattice in Fe$_{1-x}$Co$_{x}$Si may be observed as a metastable state down to the lowest temperatures studied. This behavior is most pronounced for field along $\langle100\rangle$ and weakest for $\langle111\rangle$. The dense hatching in Fig.~\ref{figure08} marks the part of the phase diagram, where the ac susceptibility stays at its lowest value, i.e., where the metastable skyrmion lattice is most pronounced. Note that no metastable skyrmion lattice was reported for Mn$_{1-x}$Fe$_{x}$Si and Mn$_{1-x}$Co$_{x}$Si.

\begin{figure}
\includegraphics[width=1.0\linewidth]{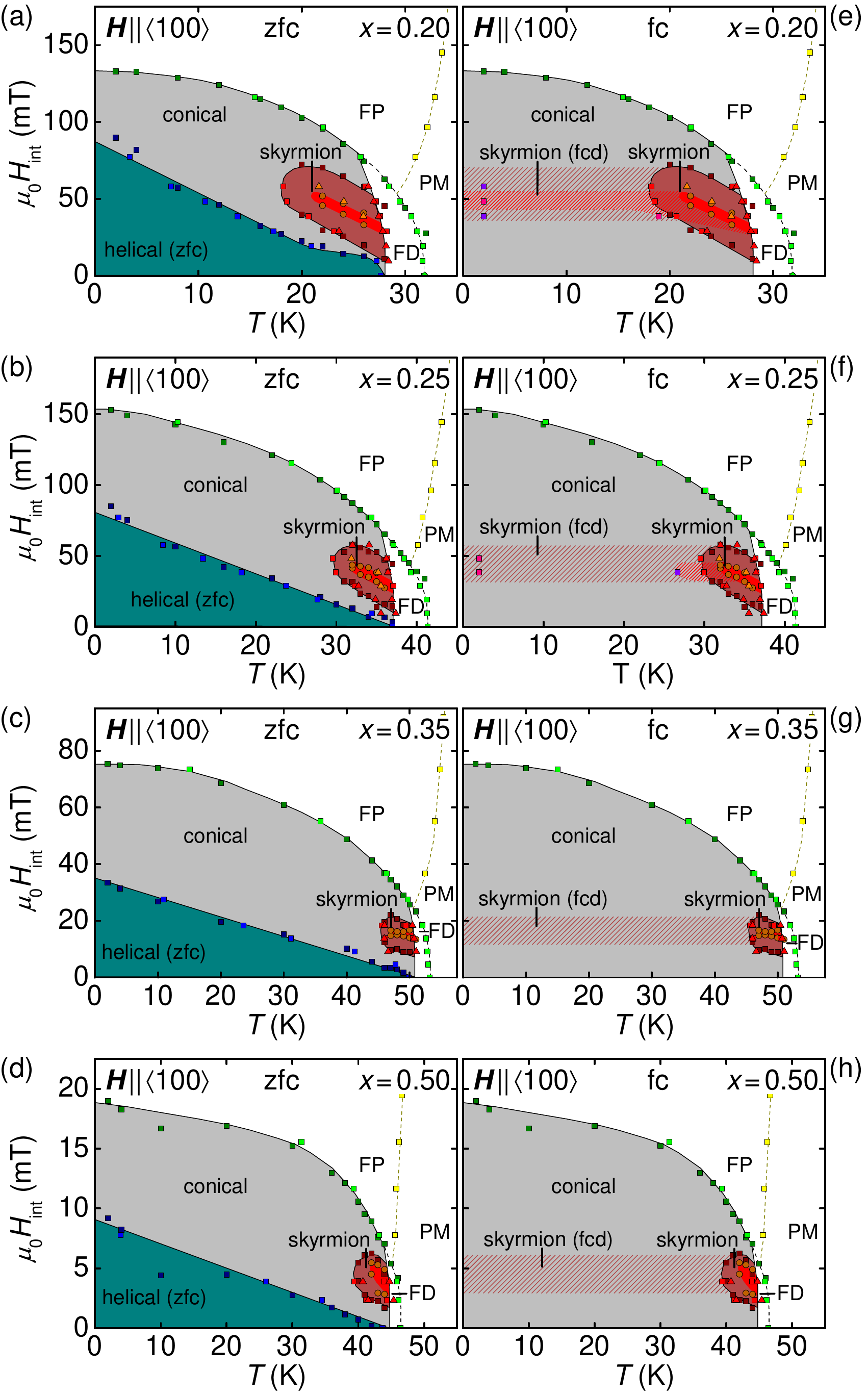}
\caption{(Color online) Magnetic phase diagram of Fe$_{1-x}$Co$_{x}$Si for various concentrations after zero-field cooling (left) and field cooling (right). The magnetic field was parallel $\langle100\rangle$. Data are shown as a function of internal field, i.e., after correcting for demagnetization effects. Depending on the cobalt content $x$ the critical temperature and field values vary, while the magnetic phase diagram stays qualitatively the same in the concentration range studied.}
\label{figure09}
\end{figure}

Figure~\ref{figure09} finally summarizes the magnetic phase diagrams for all concentrations studied. The magnetic field was applied along $\langle100\rangle$ after zero-field cooling (left column) or field cooling (right column), respectively. While the qualitative shape of all phase diagrams is very similar, the characteristic temperature and field values vary considerably as a function of cobalt content. All phase diagrams are in excellent agreement with corresponding results from small-angle neutron scattering.\cite{1995:Ishimoto:PhysicaB, 2007:Grigoriev:PhysRevB, 2010:Munzer:PhysRevB} The helimagnetic ordering temperature, $T_{c}$, reaches its maximum of more than 50\,K around $x = 0.35$, cf.\ Fig.~\ref{figure01}(a). For larger and smaller cobalt contents $T_{c}$ decreases smoothly. The critical field $H_{c2}$ peaks around $x = 0.25$ and shrinks rapidly for larger cobalt contents.

\subsection{Specific heat and entropy}
\label{SpecificHeat}

\begin{figure}
\includegraphics[width=1.0\linewidth]{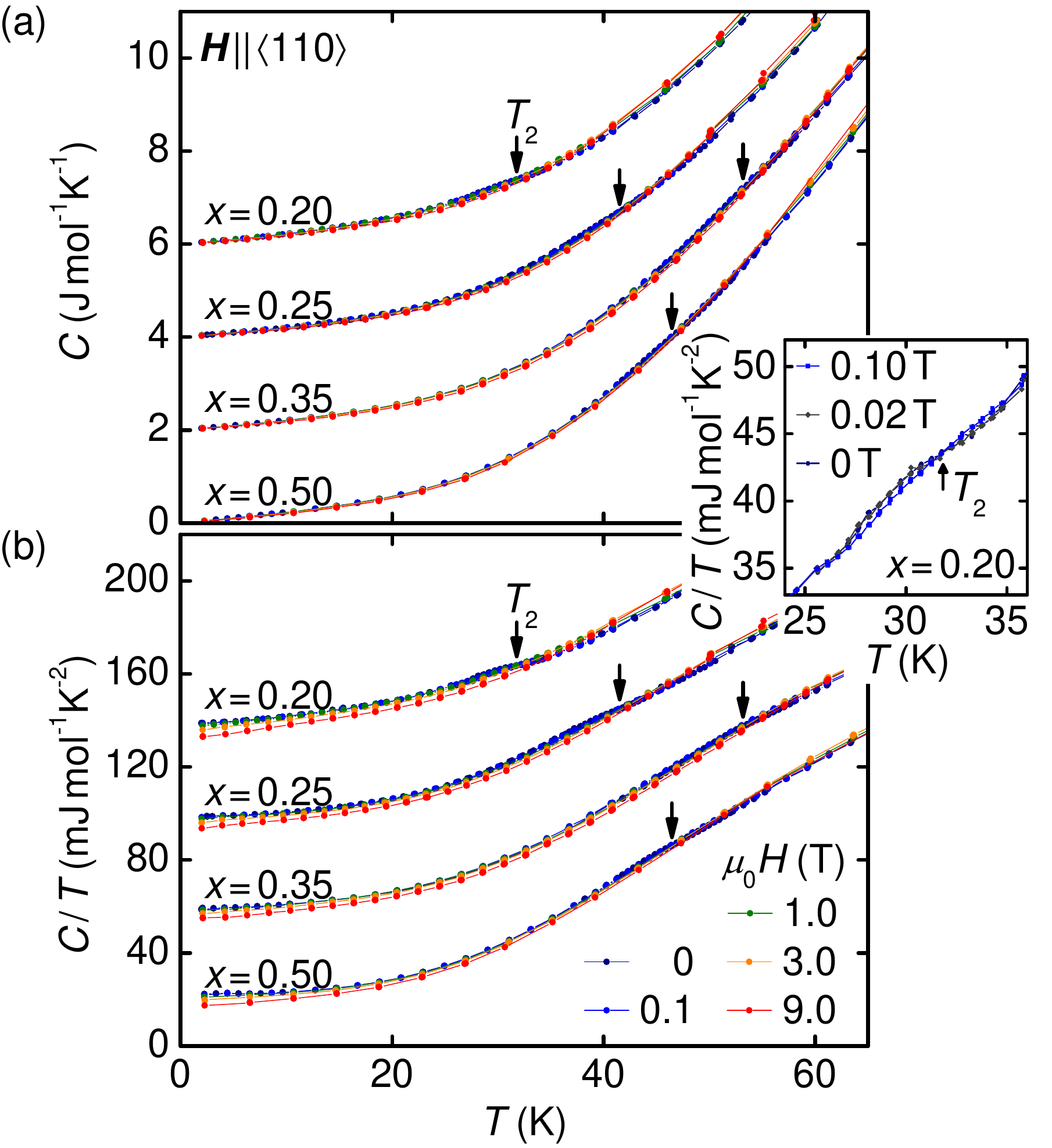}
\caption{(Color online) Temperature dependence of the specific heat of Fe$_{1-x}$Co$_{x}$Si. The magnetic field was applied along $\langle110\rangle$ after zero-field cooling. (a)~Specific heat for different cobalt contents, $x$, and magnetic fields. Data are offset by 2\,J\,mol$^{-1}$K$^{-1}$ for clarity. A Vollhardt invariance at $T_{2}$ is observed in small fields for all concentrations studied. (b)~Specific heat divided by temperature. Data are offset by 40\,mJ\,mol$^{-1}$K$^{-2}$ for clarity. The inset highlights the crossing point at $T_{2}$ in small fields for $x = 0.20$.}
\label{figure10}
\end{figure}

We will now describe the specific heat of Fe$_{1-x}$Co$_{x}$Si and analyze the contributions to the specific heat for $x = 0.20$. Figure~\ref{figure10}(a) shows the specific heat of Fe$_{1-x}$Co$_{x}$Si as a function of temperature for zero field and fields up to 9\,T. In the concentration range studied the specific heat is dominated by phonon contributions and depends only weakly on magnetic field. While previous studies of the specific heat focused on the behavior at a few Kelvin,\cite{1976:Kawarazaki:JPhysSocJpn, 1993:Lacerda:PhysicaB} in fact small additional contributions arise in the long-range modulated part of the phase diagram. These contributions are best resolved by plotting the specific heat divided by temperature, $C/T$, as a function of temperature as depicted in Fig.~\ref{figure10}(b). They are suppressed with increasing field.

A crossing point in small fields, marked $T_{2}$, shares the characteristics of a Vollhardt invariance~\cite{1997:Vollhardt:PhysRevLett, 2010:Bauer:PhysRevB, 2013:Janoschek:PhysRevB}. In the helimagnetic Brazovskii scenario, $T_{2}$ marks the crossover from the paramagnetic regime at high temperatures to a regime of strongly interacting chiral fluctuations at lower temperatures. This crossover is followed by a fluctuation-induced first-order transition into the helical state at $T_{c} < T_{2}$. A peak in the specific heat is thereby expected as a clear signature of the first-order transition that coincides with the onset of helical order. Within the resolution of our study, however, we observe no first-order-like anomaly in the specific heat of Fe$_{1-x}$Co$_{x}$Si, whereas helical order with pinned helices is detected in small-angle neutron scattering.\cite{2007:Grigoriev:PhysRevB, 2010:Munzer:PhysRevB, 2013:Milde:Science}

The lack of such a specific heat anomaly contrasts the situation in both the itinerant helimagnets MnSi\cite{2013:Bauer:PhysRevLett} and Mn$_{1-x}$Fe$_{x}$Si\cite{2010:Bauer:PhysRevB} as well as the local-moment helimagnet Cu$_{2}$OSeO$_{3}$\cite{2012:Adams:PhysRevLett}. Fe$_{1-x}$Co$_{x}$Si, although often being referred to as strongly doped semiconductor due to the temperature dependence of its electrical resistivity, behaves still as an itinerant-electron system. This assumption is corroborated by the absolute value of the resistivity, the small size of the ordered moment compared to the fluctuating moment, and the missing saturation of the magnetization up to the highest fields studied. Optical reflectivity and conductivity measurements further imply a charge carrier density that may be attributed to the intrinsic electronic structure of Fe$_{1-x}$Co$_{x}$Si rather than to an impurity band.\cite{2006:Mena:PhysRevB} Compared to pure and doped MnSi, however, the conduction electron density appears to be reduced by a factor of ${\sim}5$\cite{2004:Manyala:NatureMater} yielding comparably small magnetic contributions to the specific heat and the entropy. As a result, the specific heat anomaly could not be resolved within the resolution of our study.

\begin{figure}
\includegraphics[width=1.0\linewidth]{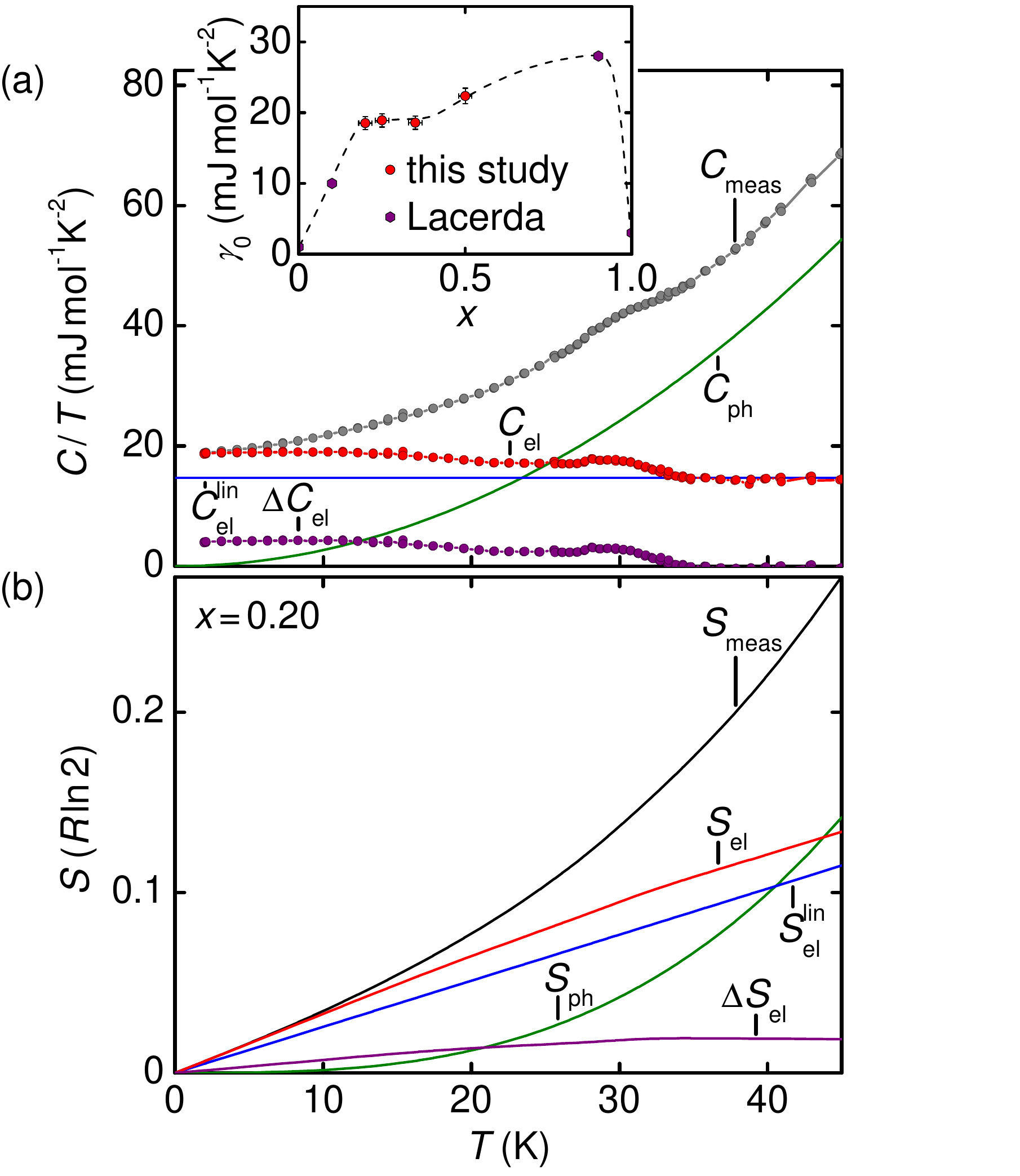}
\caption{(Color online) Contributions to the specific heat and the entropy of Fe$_{1-x}$Co$_{x}$Si with $x = 0.20$ as a function of temperature. (a)~Contributions to the specific heat divided by temperature. $C_{\mathrm{meas}}$ is the measured specific heat. $C_{\mathrm{ph}} \propto T^{3}$ is the phonon contribution according to the Debye model. $C_{\mathrm{el}} = C_{\mathrm{meas}} - C_{\mathrm{ph}}$ is the total electronic contribution. $C_{\mathrm{el}}^{\mathrm{lin}} \propto T$ is the linear electronic contribution expected from Fermi-liquid theory. $\Delta C_{\mathrm{el}} = C_{\mathrm{el}} - C_{\mathrm{el}}^{\mathrm{lin}}$ represents additional magnetic contributions. The inset shows the zero-temperature extrapolation of $C/T$, denoted $\gamma_{0}$, as a function of the cobalt content. In addition, we show data from Lacerda et al.\cite{1993:Lacerda:PhysicaB}. (b)~Contributions to the entropy as calculated from panel~(a).}
\label{figure11}
\end{figure}

An analysis of the different contributions to the specific heat supports the itinerant character of Fe$_{1-x}$Co$_{x}$Si in analogy to MnSi at reduced conduction electron density. Such an analysis is presented in Fig.~\ref{figure11}(a) for the example of $x = 0.20$ in the form of the specific heat divided by temperature, $C/T$, as a function of temperature. Here, $C_{\mathrm{meas}}$ is the measured total specific heat of the sample. $C_{\mathrm{ph}} \propto T^{3}$ corresponds to the lattice contribution derived from a Debye model using $\mathit{\Theta}_{\mathrm{D}} = 525$\,K.\footnote{By plotting $C/T$ as a function of the square of the temperature (not shown), a linear fit for $T > T_{2}$ yields a Debye temperature of $\Theta_{\mathrm{D}} = 525$\,K and an electronic high-temperature Fermi-liquid contribution $\gamma_{\mathrm{fl}} = 15\,$mJ\,mol$^{-1}$K$^{-2}$.} This value of $\mathit{\Theta}_{\mathrm{D}}$ is in good agreement with the values observed in MnSi and Mn$_{1-x}$Fe$_{x}$Si.\cite{2010:Bauer:PhysRevB} Extrapolating $C/T$ to zero temperature yields $\gamma_{0} = 19\,$mJ/mol\,K$^{2}$ for $x = 0.20$. This value is half of what is seen in MnSi and slightly increases for increasing cobalt content, $x$, in the concentration range studied, as depicted in the inset of Fig.~\ref{figure11}(a).

$C_{\mathrm{el}} = C_{\mathrm{meas}} - C_{\mathrm{ph}}$ corresponds to the total electronic contribution to the specific heat and may be split into two contributions. First, $C_{\mathrm{el}}^{\mathrm{lin}} = \gamma_{\mathrm{fl}}T$ is the linear contribution expected from Fermi liquid theory. Fitting at temperatures well above $T_{2}$, we extract $\gamma_{\mathrm{fl}} = 15\,$mJ/mol\,K$^{2}$, i.e., a value comparable to pure and doped MnSi and slightly larger than the values reported previously\cite{1976:Kawarazaki:JPhysSocJpn}. In combination with the small electron density of Fe$_{1-x}$Co$_{x}$Si, this value of $\gamma_{\mathrm{fl}}$ implies a distinct enhancement of the effective electron masses. Second, $\Delta C_{\mathrm{el}} = C_{\mathrm{el}} - C_{\mathrm{el}}^{\mathrm{lin}}$ covers the remaining specific heat. This additional magnetic contribution is tiny, which becomes especially obvious when considering the entropy calculated from the specific heat data. As shown in Fig.~\ref{figure11}(b), $\Delta C_{\mathrm{el}}$ yields an additional entropy of $\Delta S = 0.02\,R\,\ln2$. Hence, compared to MnSi and Mn$_{1-x}$Fe$_{x}$Si, this value is reduced by the same factor of 5 as the charge carrier density underscoring the itinerant character of the magnetism in Fe$_{1-x}$Co$_{x}$Si.\cite{2004:Manyala:NatureMater, 2010:Bauer:PhysRevB}


\section{Conclusions}
\label{Conclusions}

In summary, we have studied the magnetization, ac susceptibility, and specific heat of the cubic chiral magnet Fe$_{1-x}$Co$_{x}$Si for $0.20 \leq x \leq 0.50$. The magnetic properties are characteristic of an itinerant magnet with comparably low charge carrier concentration. After zero-field cooling the magnetic phase diagrams are highly reminiscent of other cubic chiral helimagnets such as MnSi or Cu$_{2}$OSeO$_{3}$, including, in particular, a single pocket of skyrmion lattice phase for all major crystallographic directions for the entire concentration range studied. Under field cooling through this pocket, however, the skyrmion lattice may persist as a metastable state down to the lowest temperatures. Moreover, no helical state is recovered once a field large enough to enter the conical state has been applied. This history dependence of Fe$_{1-x}$Co$_{x}$Si in combination with the large compositional range exhibiting helimagnetism permits to tune the temperature, field, and length scale of the magnetic order. In this context, our findings pave the grounds for studies of the interplay of disorder with the magnetic properties of Fe$_{1-x}$Co$_{x}$Si in the future.

\acknowledgments

We wish to thank T.~Adams, A.~Chacon, C.~Franz, M.~Halder, S.~Mayr, W.~M\"{u}nzer, and A.~Neubauer for fruitful discussions and assistance with the experiments. Financial support through DFG TRR80 (From Electronic Correlations to Functionality), DFG FOR960 (Quantum Phase Transitions), and ERC AdG (291079, TOPFIT) is gratefully acknowledged. A.B.\ acknowledges financial support through the TUM graduate school.

\end{document}